\documentclass[aps,prl,twocolumn,showpacs,amsmath,amssymb,floatfix]{revtex4}
\usepackage[american]{babel}
\usepackage[latin1]{inputenc}
\usepackage{amsmath}
\usepackage{amssymb}
\usepackage{bm}
\usepackage{epsfig}
\usepackage{graphicx}
\usepackage{euscript}
\usepackage{color}
\usepackage{float}
\usepackage{framed}
\usepackage{mathrsfs}
\usepackage{hyperref}
\usepackage{multirow}
\usepackage{subeqnarray}
\usepackage{bbold}
\usepackage{upgreek}
\usepackage{nicefrac}

\begin{document}

\title{Pair creation in heavy ion channeling}


\author{N.~A.~Belov}
\author{Z.~Harman}
\affiliation{Max Planck Institute for Nuclear Physics, Saupfercheckweg 1, 69117 Heidelberg, Germany}



\begin{abstract}

Heavy ions channeling through crystals with multi-GeV kinetic energies can create electron-positron pairs.
In the framework of the ion, the energy of virtual photons arising from the periodic
crystal potential may exceed the threshold $2mc^2$. The repeated periodic collisions with the crystal ions
yield high pair production rates. When the virtual photon frequency matches a nuclear transition in the ion, the production rate
can be resonantly increased. In this two-step excitation-pair conversion scheme, the excitation rates
are coherently enhanced, and they scale approximately quadratically with the number of crystal sites along the channel.

\end{abstract}

\maketitle

The creation of particle-antiparticle pairs from vacuum~\cite{Dir28} is one of the most intriguing features of quantum field theory.
A broad variety of pair creation (PC) mechanisms have been predicted and experimentally observed,
e.g., in intense optical or x-ray fields~\cite{Sar13,Gon13,DiP12,ELI,Che10,Fed10,Ruf09,DiP09,Bel08,Sch08,Mul03,Alk01},
in ion-ion~\cite{Bau07,Li98,Van92,Rum91,Gre85}, ion-photon~\cite{CSO97} and ion-electron~\cite{Art03} collisions,
in tokamak plasmas~\cite{Hel03}, as well as in astrophysical environments such as, e.g., pulsars~\cite{Sha82}.
In projected experiments at the Facility for Antiproton and Ion Research (FAIR), it will be possible to study PC in the Coulomb field of
heavy ions during their collision~\cite{FAIR-CDR,FAIR}.

\begin{figure}[t]
   \center
   \includegraphics[width=0.9 \columnwidth]{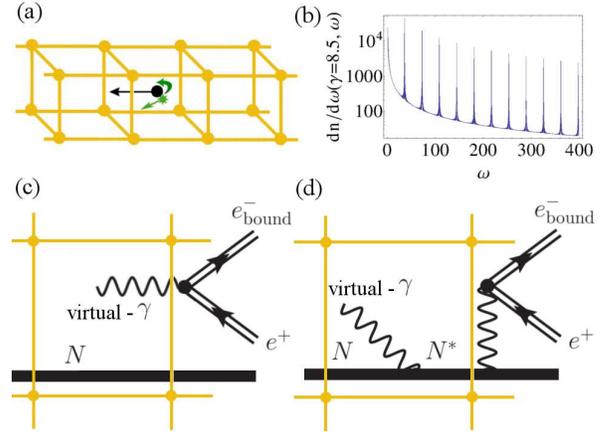}
   \caption{(a) Schematic view of pair production in heavy ion channeling. (b) Typical virtual photon spectrum of the crystal.
   Diagrams for (c) direct production by an equivalent photon and (d) pair creation proceeding through nuclear excitation.}
   \label{fig:Chan_common}
\end{figure}

In this Letter we put forward an alternative mechanism of electron-positron PC
in ion planar channeling through a crystal [see Fig.~\ref{fig:Chan_common}(a)].
In the reference frame of the travelling ions, the electromagnetic field of the periodic crystal structure may be regarded as a
field of virtual photons with well-defined, equidistantly spaced discrete frequencies. For fast ions, these frequencies may
extend into the MeV range, surpassing the PC threshold. In a direct channeling PC process, at all photon energies
above this threshold value, a free-free or bound-free pair can be created; in the latter case, the electron is immediately
captured into a bound state of the ion. In addition, when the virtual-photon frequency matches a nuclear transition
in the channeling ion, a two-step resonant process may occur, in which first the nucleus is excited, then it decays
by internal pair conversion. After multiple interfering periodic interactions of the channeling heavy ion with the crystal
sites, PC occurs with significantly enhanced probability as compared to the collision of single ions. PC with
channeling ions may also be regarded as a feasible alternative to photo-production with a currently non-existing intense
coherent gamma-ray field.

Atomic and nuclear resonant excitations in axial channeling were firstly described by
Okorokov~\cite{Oko65, Oko65_2}. Recently, resonant coherent excitation (RCE) of ions was experimentally
investigated~\cite{Nak13,Nak09,Nak08,Tes07,Kon06,And97,And92}, with ions as heavy as ${}^{238}$U${}^{89+}$~\cite{Nak13}
and transition energies as high as 6.7~keV~\cite{Nak09}. These experiments are planned to be extended to the 100-keV
excitation of hydrogenlike ${}^{238}$U${}^{91+}$~\cite{Sto14,Nak13}, therefore, it is reasonable to anticipate that
further developments will reach the MeV regime of PC.
A general formalism for atomic excitations is presented in, e.g., Refs.~\cite{Bal06,Shi76,Cra79}, while in
Refs.~\cite{Piv86,Piv90}, a framework suitable for describing nuclear excitation in channeling has been developed.
Crystal-assisted PC by synchrotron radiation gamma photons has been theoretically formulated
in~\cite{Cue85}. In this process, only free leptons can be produced, thus one cannot exploit the advantage of
resonances of the virtual-photon density of the crystal.

The two cases of direct and nuclear-resonant PC are schematically presented in Fig.~\ref{fig:Chan_common}(c) and (d), respectively.
Electromagnetic processes in ion channeling can be described by the presence of virtual photons of the crystal field (see, e.g., \cite{Piv90}).
The spectral density of virtual (equivalent) photons of frequency $\omega$ can be
derived~\cite{Piv86} by the help of the classical Weizs\"acker-Williams method~\cite{Jac99}. This approximation is valid in the
ultrarelativistic case, i.e. when the Lorentz factor $\gamma=\left(1-\nicefrac{v^2}{c^2}\right)^{-1/2} \gg 1$ (with $v$ being is the ion velocity
and $c$ the speed of light), and yields the spectral density
\begin{eqnarray}
\frac{dn(\gamma,\omega)}{d\omega}=\frac{I_2(\omega)}{v \hbar\omega(2\pi)^4}
\frac{\sin^2\left(\frac{\omega aN}{2\gamma v}\right)}{\sin^2\left(\frac{\omega a}{2\gamma v}\right)}
e^{-\left(\frac{\omega\delta}{\gamma v}\right)^2} \nonumber \\
+\frac{I_2(\omega)}{v \hbar\omega(2\pi)^4}
N\left(\frac{I_1(\omega)}{I_2(\omega)}-e^{-\left(\frac{\omega\delta}{\gamma v}\right)^2}\right)\,.
\label{eqn:Chan_VP}
\end{eqnarray}
Here, $\hbar$ stands for the reduced Planck constant, $N$ is the number of atoms in a crystal channel, $a$ denotes the lattice
constant, and $\delta$ stands for the amplitude of thermal oscillations. The integrals $I_1$ and $I_2$ are given by
\begin{eqnarray}
I_1(\omega)&=&\int d^2{\bf k}_\perp {\bf k}_\perp^2V_{\bf k}^2\,, \nonumber \\
I_2(\omega)&=&\int d^2{\bf k}_\perp {\bf k}_\perp^2V_{\bf k}^2 \exp^{-{\bf k}_\perp^2\delta^2}\,,
\end{eqnarray}
with the 2-dimensional transverse wave vector
$ {\bf k}_\perp^2={\bf k}^2-\left(\frac{\omega}{\gamma v}\right)^2$, and $V_{\bf k}$ being the Fourier transform of a
single atom's potential in the crystal.
It follows from Eq.~(\ref{eqn:Chan_VP}) that at the energies $\omega_n=2\pi n\gamma v/a$, $n \in \left[1, 2, 3, \dots \right]$,
the virtual photon spectrum exhibits maxima proportional to $N^2$ [see Fig. \ref{fig:Chan_common}(b)].
The photon energies $\omega_n$ can be experimentally tuned by choosing the proper $\gamma$. Due to restrictions
caused by thermal vibrations of the lattice atoms, typically harmonics with $n < 10$ are used~\cite{Piv96}.

{\it Direct PC process.}
In this process, illustrated in Fig.~\ref{fig:Chan_common}(c), the outgoing positrons possess a continuous spectrum for free-free PC,
and a monochromatic energy in the bound-free case.
We define, following~\cite{Piv86}, the cross section $\sigma_{\rm PC}^{\rm chan}$ of PC via channeling as the convolution of
the virtual photon density with the cross section $\sigma_{\rm PC}$ of PC by a real photon:
\begin{equation}
\sigma^{\rm chan}_{\rm PC}(\gamma)=\int d\omega\sigma_{\rm PC}(\omega)\frac{dn(\gamma,\omega)}{d\omega}\,,
\label{eqn:Chan_main}
\end{equation}
where $\sigma_{\rm PC}(\omega)$ can either represent the cross section of bound-free (bf) or free-free (ff) PC.
The number of pairs created in unit time can be expressed as $\dot{N}^{\rm chan}_{\rm PC}=S \Phi {\sigma^{\rm chan}_{\rm PC}}/{a^2}$,
with $S$ being the cross sectional area of the ion beam and $\Phi$ its flux.
The cross section for bound-free PC in the Coulomb nuclear field by an external photon is~\cite{CSO97}
\begin{equation}
\sigma_{\rm PC,bf}(\omega) =
\frac{2\pi^2\alpha\lambda_C^2}{\omega}\sum_{\lambda}\sum_{JLM M_{\rm b}} \left|{\mathcal M}_{\lambda JLM M_{\rm b}}(\omega)\right|^2 \,,
\end{equation}
where $\alpha$ is the fine-structure constant, $\lambda_C$ denotes the electrons Compton wavelength, $\lambda=\pm 1$ is the helicity of the photon,
$M_{\rm b}$ is the magnetic quantum number of the created bound electron, and the quantum numbers $J$, $L$ and $M$ correspond
to the partial waves of the free-positron wave function. The matrix element is given as
\begin{equation}
{\mathcal M}_{\lambda M_{\rm b} JLM} (\omega) =
\int d^3{\bf r}\psi_{\rm b}^*({\bf r}) \left[{\boldsymbol \alpha} \cdot {\bf e}_\lambda\right] e^{i\omega r}\psi_{JLM}({\bf r})\,,
\end{equation}
where $\psi_{JLM}$ is the positron wave function in the Coulomb field of the nucleus, $\psi_{\rm b}$ is the bound-electron wave function,
and ${\boldsymbol \alpha}$ is the 3-vector of alpha matrices. Formulas for free-free PC may be similarly derived,
involving an additional summation over the partial waves of the free electron. This approach, neglecting the interaction
of the created particles with the periodic field, can be employed because, for the high frequencies present here,
the classical nonlinearity parameter $\xi_0 = \frac{eE}{m\omega}$~\cite{DiP12}, written in terms of the crystals electric
field strength $E$ in the framework of the ion and the unit charge $e$, is much less than unity. $\sigma_{\rm PC,bf}$ has an
energy threshold at $2mc^2-E_{\rm b}$, with $E_{\rm b}$ being the binding energy of the created electron. After this threshold,
$\sigma_{\rm PC,bf}$ increases with energy up to a given maximal value, whose position and value depend on the nuclear charge.

In Tab.~\ref{tab:Chan_direct} we present cross-section values for direct PC for different charges $Z$ of the ion channeling in an Au
crystal, at energy $\omega'$, corresponding to the maximum of the direct bound-free PC cross section. We choose to match the energy
of the 6th harmonic of the virtual photon density~(\ref{eqn:Chan_VP}). The advantage of using heavy ions is justified by the
following scaling low: the cross section $\sigma_{\rm PC}$ scales with the charge number as $ \approx Z^{5-\epsilon}$, $0 \le \epsilon \le 1$,
and the $Z$-scaling of $\sigma^{\rm chan}_{\rm PC}$ is given by that of $\sigma_{\rm PC}$ [see Eq.~(\ref{eqn:Chan_main})].
At high virtual photon energies, the PC cross section decreases and vanishes asymptotically.
The cross section $\sigma^{\rm chan}_{\rm PC}$ is compared to the cross section $\sigma^{\rm coll}_{\rm PC}$ of the PC
in the single Coulomb collision process (estimated from the channeling cross section with the substitution of $N=1$),
and with the cross section $\sigma_{\rm PC}$ of pair photo-production in the nuclear Coulomb field~\cite{CSO97}.
In both cases, one can see a significant increase. The cross sections can be translated to pair creation rates, assuming e.g.
an ion beam cross section of $S=1$~cm$^2$ and flux $\Phi=10^{10}/({\rm cm}^2 {\rm s})$, yielding $\dot{N}^{\rm chan}_{\rm PC,bf}=$600/s,
2900/s and 7800/s, for $Z=$50, 75 and 92, respectively.

\begin{table}
\begin{center}
\caption{\label{tab:Chan_direct}
Cross sections (in barn) for the direct bound-free/free-free
pair production by a bare ion with a charge $Z$, and $\gamma$ corresponding to the energy $\omega'$ for $n=6$ and $N=100$.
$\omega'$ (in MeV) is the maximum of the direct bound-free pair production cross section, following Ref.~\cite{CSO97}.
The notation $a[b]$ stands for $a \times 10^b$.
}
\begin{tabular}{cccccc}
  \hline
  \hline
    $Z$ & $\omega'$ & $\gamma$ & $\sigma^{\rm chan}_{\rm PC}$ & $\sigma^{\rm coll}_{\rm PC}$ & $\sigma_{\rm PC}$ \\
  \hline
    1  & 3.2 & 123  & $8.3[-6]$/$1.1[3]$ & $8.3[-8]$/$1.2[1]$ & $2.8[-10]$/$7.6[-10]$ \\
    25 & 3.1 & 119  & $4.5[1]$/$7.0[5]$  & $4.5[-1]$/$7.2[3]$ & $1.3[-3]$/$3.4[-3]$  \\
    50 & 3.1 & 119  & $1.0[3]$/$2.8[6]$  & $1.0[1]$/$2.9[4]$  & $2.6[-2]$/$6.8[-2]$  \\
    75 & 1.5 & 59.3 & $4.9[3]$/$5.7[6]$  & $4.9[1]$/$5.8[4]$  & $1.7[-1]$/$7.5[-2]$  \\
    92 & 1.5 & 59.3 & $1.3[4]$/$8.6[6]$  & $1.3[2]$/$8.7[4]$  & $4.6[-1]$/$2.1[-1]$ \\
  \hline
  \hline
\end{tabular}
\end{center}
\end{table}

One may define a ratio $R(\gamma)$ of the cross section of the coherent interaction with $N$ atoms to the incoherent one
as follows: $R(\gamma)=\sigma^{\rm chan}_{\rm PC}(\gamma)/\left( N \sigma^{\rm coll}_{\rm PC}(\gamma)\right)$.
This parameter measures the coherence in the channeling process. For the cases shown in Tab.~\ref{tab:Chan_direct},
$R$ is practically equal to unity, showing that there is no enhancement due to coherence for direct PC via channeling.
This is explained by the broadness of the continuous spectrum of photons which can create a pair, as compared to the width
$2\gamma v/(aN)$ of a photon density peak. In other words, the coherence length of photo-production is much shorter
than the Lorentz-contracted crystal length $aN/\gamma$.

High ion kinetic energies are needed to meet the resonance conditions for the virtual-photon energies
required for direct PC by ion channeling. Such energies can be reached, for instance,
by the projected FAIR accelerators~\cite{FAIR} or in the Large Hadron Collider~\cite{LHC}. Laser-accelerated
ion beams, anticipated to eventually reach the GeV regime~\cite{Dai12,Esi04}, may also provide a viable alternative in future.
One possible way to reduce the required ion kinetic energy is to use higher crystal-field harmonics.
This may be more feasible at high ion energies than at the low energies of the RCE experiments performed thus
far~\cite{Nak13,Nak09,Nak08,Tes07,Kon06,And97,And92}, because fast ions interact less with the crystal electrons,
suppressing decoherence.

{\it PC proceeding via nuclear resonances.}
The cross section of the first step of the process shown on Fig.~\ref{fig:Chan_common}(d), namely, the RCE of the nucleus
passing through the crystal, can be given as~\cite{Piv86,Piv90}
\begin{equation}
\sigma^{\rm chan}_{\rm N}(\gamma)=\int d\omega \sigma_{\rm N}(\omega)\frac{dn(\gamma,\omega)}{d\omega}\,,
\end{equation}
with the cross section of nuclear excitation with a real photon of frequency $\omega$,
\begin{equation}
\sigma_{\rm N}(\omega)=g\frac{\pi c^2}{\omega_0^2}\frac{\Gamma_{\rm rad}^2}{(\omega-\omega_0)^2+\Gamma^2/4}\,,
\end{equation}
where $\omega_0$ is the nuclear excitation energy. The statistical factor $g=(2I_f+1)(2I_i+1)$ depends on the angular momenta $I_i$ ($I_f$)
of the initial (final) nuclear states, $\Gamma$ is the total width of the excited nuclear level, and $\Gamma_{\rm rad}$ is its radiative width.
To obtain the cross section for the total two-step process of ion excitation--de-excitation by PC, $\sigma^{\rm chan}_{\rm N}$ is multiplied
by the coefficient of pair conversion. We introduce this coefficient $\beta_{\rm bf}$ for the bound-free case to be the ratio of the
transition probabilities of PC and radiative decay: $\beta_{\rm bf}={P_{\rm bf}}/{P_{\rm rad}}$. This rate can be calculated, e.g.,
following Refs.~\cite{OUR_NERPA_letter,SSG81_1}. One obtains the following expressions in dependence of the corresponding nuclear transition
multipolarity, i.e., the angular momentum $L'$ and parity of the transition:
\begin{eqnarray}
\label{eq:betabf}
&&\beta_{\rm bf}(EL')=\sum_{\kappa\kappa'}\frac{4\pi\alpha\omega}{L'(L'+1)} s |\kappa\kappa'|\\
&&\times|(\kappa-\kappa')(R_3+R_4)+L'(R_1+R_2+R_3-R_4)|^{2}\,, \nonumber\\
&&\beta_{\rm bf}(ML')=\sum_{\kappa\kappa'}\frac{4\pi\alpha\omega}{L'(L'+1)} s |\kappa\kappa'| |(\kappa+\kappa')(R_3+R_4)|^{2}\,, \nonumber
\end{eqnarray}
where we introduced in terms of a 3$j$-symbol
\begin{eqnarray}
s &=&\left(%
\begin{array}{ccc}
  j & j' & L \\
  \frac{1}{2} & -\frac{1}{2} & 0 \\
\end{array}%
\right)^{2}\,.
\end{eqnarray}
Here, $\kappa$ and $\kappa'$ are Dirac angular momentum quantum numbers.
The radial integrals $R_1, \dots ,R_6$ are defined as in~\cite{SSG81_1} with the analytical
form of the fermionic Coulomb wave functions for bound and free particles.
All results are obtained for the $1s$ electron orbital having a maximal overlap with the nucleus.
Equation~(\ref{eq:betabf}) can be adopted to the free-free case ($\beta_{\rm ff}$) in a straightforward manner.

\begin{figure}[t!]
\includegraphics[width=0.8 \columnwidth]{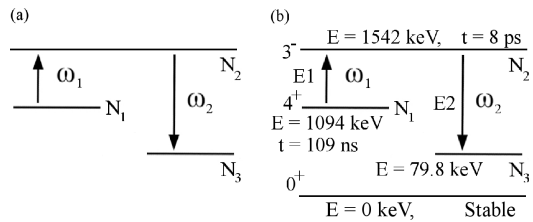}
\caption{\label{fig:Chan_levels}
(a) The preferred 3-level scheme and (b) the level scheme for one of the possible elements, $^{168}$Er.}
\end{figure}

\setlength{\tabcolsep}{3mm}
\begin{table}
\begin{center}
\caption{\label{tab:Chan_data}
Nuclear data for different elements corresponding to the level scheme introduced at Fig.~\ref{fig:Chan_levels}(a).
Energies are given in units of keV.}
\begin{tabular}{llll}
  \hline
  \hline
                   & $^{168}$Er & $^{72}$Ge & $^{115}$Sn\\
\hline
    $E_{N_1}$      & 1094       & 691.4     & 612.8\\
    $E_{N_2}$      & 1542       & 1464      & 1416.9\\
    $E_{N_3}$      & 79.8       & 0         & 0\\
    $\omega_1$     & 447.6      & 772.6     & 807.1\\
    $\omega_2$     & 1462       & 1464      & 1416.9\\
    $\tau_1$       & 109 ns     & 444 ns    & 3.26 $\mu$s\\
    $\Gamma$ (meV) & 0.082      & 0.15      & 1.88\\
$\rho_{N_2\rightarrow N_3}$ & 0.0058 & 0.124 & 0.739 \\
$\beta_{N_2\rightarrow N_3}^{\rm bf}\cdot 10^{4}$ & 9.3 & 0.55 & 2.0  \\
$\beta_{N_2\rightarrow N_3}^{\rm ff}\cdot 10^{4}$ & 0.77 & 0.74  & 0.52 \\
  \hline
  \hline
\end{tabular}
\end{center}
\end{table}

The cross section of the two-step nuclear excitation-pair conversion (NEPC) process for channeling ions can be written as
\begin{equation}
\sigma^{\rm chan}_{\rm NEPC}(\gamma)=\sigma^{\rm chan}_{\rm N}(\gamma) B \beta\,,
\label{eqn:Chan_sigma_tot}
\end{equation}
where $B$ is the branching ratio of the gamma decay corresponding to the PC transition, and $\beta=\beta_{\rm bf}$
or $\beta_{\rm ff}$. One can see in Eq.~(\ref{eqn:Chan_sigma_tot}) that the cross section for the excitation
$\sigma^{\rm chan}_{\rm N}$ and the rate for the de-excitation $\beta$ are independent and, in principle, may
correspond to different transitions to/from some excited nuclear state $N_2$. Let us rewrite the total cross section as
\begin{eqnarray}
\label{eqn:Chan_sigma_2step}
&&\sigma^{\rm chan}_{\rm NEPC} (\omega_1^{N_1\rightarrow N_2},\omega_2^{N_2\rightarrow N_3})= \\
&&\sigma^{\rm chan}_{\rm N}(\omega_1^{N_1\rightarrow N_2}) B^{N_2\rightarrow N_3}
\beta(\omega_2^{N_2\rightarrow N_3})\,. \nonumber
\end{eqnarray}
Levels $N_1$, $N_2$ and $N_3$ are depicted on the three-level scheme of Fig.~\ref{fig:Chan_levels}(a).
The energy $\omega_2^{N_2\rightarrow N_3}$ has to exceed the bound-free PC threshold $2mc^2-E_{e^-_{\rm bound}}$.
The excitation energy $\omega_1^{N_1\rightarrow N_2}$ is not restricted, and from an experimental point of view
it is preferable to utilize a transition with a lower energy. It is more advantageous for level $N_1$
to be metastable, in order to be able to prepare the nuclei in this state before injecting them into the crystal.
One of the possible elements is $^{168}$Er, with its level scheme shown in Fig.~\ref{fig:Chan_levels}(b).
Results for the pair conversion coefficients $\beta$ for this isotope, together with data for other potentially
suitable elements, $^{72}$Ge and $^{115}$Sn, are given in Tab.~\ref{tab:Chan_data}.

\begin{figure}[t]
\includegraphics[width=0.8 \columnwidth]{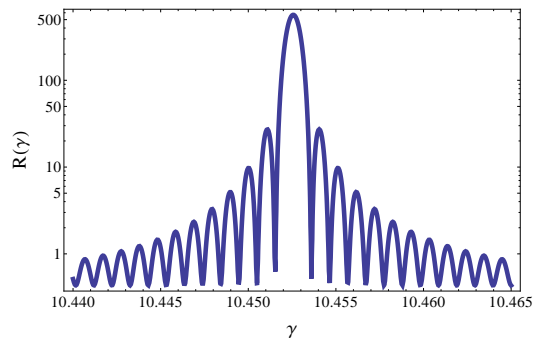}
\caption{\label{fig:Chan_R_fun} The ratio $R(\gamma)$
for the case of 447.6~keV transition in $^{168}$Er nucleus. Here, $N$=1000, and $n$=10.}
\end{figure}

Once the nuclear transitions involved and the type of crystal are fixed, the only variable parameters
are the thickness of the crystal determined by $N$, and the harmonic order $n$.
The dependence of the ion kinetic energy, connected with $\gamma$, and of the excitation cross section $\sigma^{\rm chan}_{\rm N}$ on
$n$ at certain value of $N$ is presented in Tab.~\ref{tab:Chan_N100,ndep}. The
largest cross section is reached at the fundamental frequency, however, the cross section
decreases slowly with increasing $n$, therefore, it is again preferable to tune the $\gamma$ factor to higher harmonics.
Values of the coherent enhancement factor $R$ are also given in Tab.~\ref{tab:Chan_N100,ndep}.
$R(\gamma)$ only weakly depends on the element and the transition, and is mostly influenced by the harmonic order $n$, and by $N$.
The dependence of $R$ on $\gamma$ is shown in Fig.~\ref{fig:Chan_R_fun} for the case of the 447.6-keV transition in $^{168}$Er.
The figure shows that, for a crystal as thin as 1000 atoms, which is typically used in experiments~\cite{Nak13},
one can achieve a coherent PC enhancement by 3 orders of magnitude. By the help of high-energy ion beams
(up to 33~GeV/u or $\gamma$=35) provided by the FAIR facility in the near future~\cite{FAIR}, one can
investigate all elements in Tab.~\ref{tab:Chan_N100,ndep}.

\setlength{\tabcolsep}{2mm}
\begin{table}
\begin{center}
\caption{\label{tab:Chan_N100,ndep} The dependence of the ion's $\gamma$ factor
and the excitation cross section $\sigma^{\rm chan}_{\rm N}$ (in barn) on the harmonic order $n$ at certain value of $N=100$
for different $\omega_1$. The last column gives the ratio $R$, depending on $n$ only and not on the atomic properties
for all narrow transitions (i.e. those with a line width below the bandwidth of the virtual photon spectrum).}
\begin{tabular}{cccccccc}
  \hline
  \hline
            & \multicolumn{2}{c}{${}^{168}$Er} & \multicolumn{2}{c}{${}^{72}$Ge} & \multicolumn{2}{c}{${}^{115}$Sn} & \\
$\omega_1$= & \multicolumn{2}{c}{447.6~keV}    & \multicolumn{2}{c}{772.6~keV}   & \multicolumn{2}{c}{807.1~keV}    & \\
  $n$       & $\gamma$ & $\sigma^{\rm chan}_{\rm N}$ & $\gamma$ & $\sigma^{\rm chan}_{\rm N}$ & $\gamma$ & $\sigma^{\rm chan}_{\rm N}$ & $R$ \\
  \hline
    1  & 104& 8.2 & 180& 0.023 & 188 & 1.8 & 69 \\
    2  & 52 & 7.2 & 90 & 0.020 & 94  & 1.6 & 66 \\
    4  & 26 & 6.2 & 45 & 0.018 & 47  & 1.3 & 63 \\
    6  & 17 & 5.7 & 30 & 0.016 & 31  & 1.2 & 61 \\
    8  & 13 & 5.3 & 22 & 0.015 & 23  & 1.1 & 59 \\
   10  & 10 & 5.0 & 18 & 0.014 & 19  & 1.1 & 57 \\
  \hline
  \hline
\end{tabular}
\end{center}
\end{table}

The first two lines of the Tab.~\ref{tab:Chan_tot_sigma} demonstrate the behavior of the nuclear excitation cross section
$\sigma^{\rm chan}_{\rm N}$ on $N$, the number of ion sites along the channel. One can observe a significant -- quadratic --
enhancement of the cross section with increase of $N$. Employing a thicker crystal with higher $N$ is experimentally limited by
restrictions due to deviations from a straight ion trajectory in the crystal. One may provide more precise Monte-Carlo simulations
to model the ion trajectories, if it is necessary to increase the crystal thickness even further.
The total cross section of the two-step process NEPC can be calculated by Eq.~(\ref{eqn:Chan_sigma_2step}), using data from
Tables~\ref{tab:Chan_data} and~\ref{tab:Chan_N100,ndep}. In Tab.~\ref{tab:Chan_tot_sigma} (last 4 rows) results are presented
for this cross section for different thicknesses of the crystal $(N=100$, $1000)$ with a fixed harmonic number $(n=6)$. The
rate of created pairs is also proportional to $N^2$.

\setlength{\tabcolsep}{3mm}
\begin{table}
\begin{center}
\caption{\label{tab:Chan_tot_sigma} 
The cross section (in barn) of bound-free and free-free pair creation via nuclear resonances
and the coherence parameter $R$, for different crystal thicknesses $(N=100,~1000)$ with a fixed harmonic order $(n=6)$.
}
\begin{tabular}{lrrrr}
  \hline
  \hline
              &   $N$ & $^{168}$Er & $^{72}$Ge & $^{115}$Sn\\
\hline
$\sigma_{\rm N}$ & 100  & 5.7[0]  & 1.6[-2] & 1.2[0] \\
                 & 1000 & 5.6[2]  & 1.6[0]  & 1.2[2]  \\
$R(\gamma)$  & 100  & 60.5 & 60.6   & 60.5 \\
             & 1000 & 601  & 604    & 602  \\
$\sigma_{tot}^{\rm bf}$  & 100  & 3.1[-5] & 1.1[-7] & 1.8[-4] \\
                         & 1000 & 3.0[-3] & 1.1[-5] & 1.8[-2] \\
$\sigma_{tot}^{\rm ff}$  & 100  & 2.5[-6] & 1.4[-7] & 4.7[-5] \\
                         & 1000 & 2.5[-4] & 1.4[-5] & 4.6[-3] \\
  \hline
  \hline
\end{tabular}
\end{center}
\end{table}

{\it Summary.}
Direct PC and PC via nuclear excitation in heavy ion collisions can be significantly enhanced by multiple periodic collisions
in a crystal channeling experiment. The direct process is associated with high creation rates increasing approximately linearly
with the number of crystal sites in the channel. As for PC proceeding via nuclear resonances which typically have lower probabilities,
the coherent nature of the excitation process yields a quadratic scaling with the number of collisions, resulting in observable PC rates.
These studies complement PC by different strong electromagnetic fields such as optical or x-ray lasers~\cite{ELI}.
The ion kinetic energies required for such investigations can be reached by present and upcoming experimental facilities,
such as, e.g., FAIR. Laser-accelerated ions may also be considered in future for such studies.

We acknowledge insightful conversations with Christoph H. Keitel, Antonino Di Piazza and Karen Z. Hatsagortsyan.

\end{document}